\documentstyle[epsfig,longtable]{aipproc}

\begin{document}

\title{Science with the Constellation-X Observatory}  
 
\author{Azita Valinia$^{1,2}$, Nicholas White$^{1}$,
Harvey Tananbaum$^3$, and the Constellation-X Team$^4$}
\address{$^1$NASA's Goddard Space Flight Center, Code 662, Greenbelt, MD 20771\\
$^2$Department of Astronomy, University of Maryland, College Park, MD 20742\\
$^3$Harvard-Smithsonian Center for Astrophysics, 60 Garden St, Cambridge, MA 
02138\\
$^4$see http://constellation.gsfc.nasa.gov}

\maketitle

\begin{abstract}
The Constellation X-ray Mission is a high throughput X-ray facility
emphasizing observations at high spectral resolution 
(E/$\Delta$E $\sim$ 300--3000), and broad energy bandpass (0.25--40 keV). 
Constellation-X will provide a factor of nearly 100 increase in 
sensitivity over current high resolution X-ray spectroscopy missions.
It is the X-ray astronomy equivalent of large ground-based optical
telescopes such as the Keck Observatory and the ESO Very Large Telescope.
When observations commence toward the end of next decade, Constellation-X
will address many fundamental astrophysics questions such as: 
the formation and evolution of
clusters of galaxies; constraining the baryon content of the Universe;
determining the spin and mass of supermassive black holes in AGN; and 
probing strong gravity in the vicinity of black holes. 
\end{abstract}

\section*{Constellation-X}

The prime objective of Constellation-X mission is high resolution X-ray
spectroscopy. It will cover the $0.25-40$~keV X-ray bandpass by utilizing  
two types of high throughput telescope systems to simultaneously cover the 
low (0.25 to 10 keV) and high energy (6 to 40 keV) bands. 
The low-energy Spectroscopy
X-ray Telescope (SXT) is optimized to maintain a spectral resolving power
of at least 300 across the 0.25 to 10 keV band pass (E/$\Delta$E $\sim$ 3000 
at 6~keV) and has a minimum
telescope angular resolution of $15''$ HPD. The diameter of the field of
view is $2.5'$ below 10~keV. The high energy system
(HXT) with lower spectral resolving power ($\Delta$E $\sim$ 1~keV) overlaps 
the SXT and primarily
is used to measure the relatively line-less continuum from 10 to 40 keV. 
The diameter of the field of view is $8'$ for the HXT. 
The large collecting area is achieved with a design utilizing several
mirror modules, each with its own spectrometer/detector system. 
The spectral resolving power of the SXT 
and the effective area of SXT and HXT are
shown in Figure~1.  

The SXT uses two spectrometer systems that operate simultaneously to 
achieve the desired energy resolution: 1) a 2 eV resolution quantum
microcalorimeter array, and 2) a set of reflection gratings for 
energies $< 2$ keV. The gratings deflect part of the telescope beam
away from the calorimeter array in a design similar to XMM except that
the direct beam falls on a quantum calorimeter instead of on a CCD. 
The two spectrometers are complementary, with the gratings optimal for
high resolution spectroscopy at low energies and the calorimeter
at high energies. The gratings also provide coverage in the 0.3-0.5 keV
band where the calorimeter thermal and light-blocking filters cause a loss
of response. This low-energy capability is particularly important for
high-redshift objects, for which line-rich regions will be moved into
this low energy band. 
\begin{figure}[t] 
\hskip-5truein
\vskip+1truein
\epsfig{file=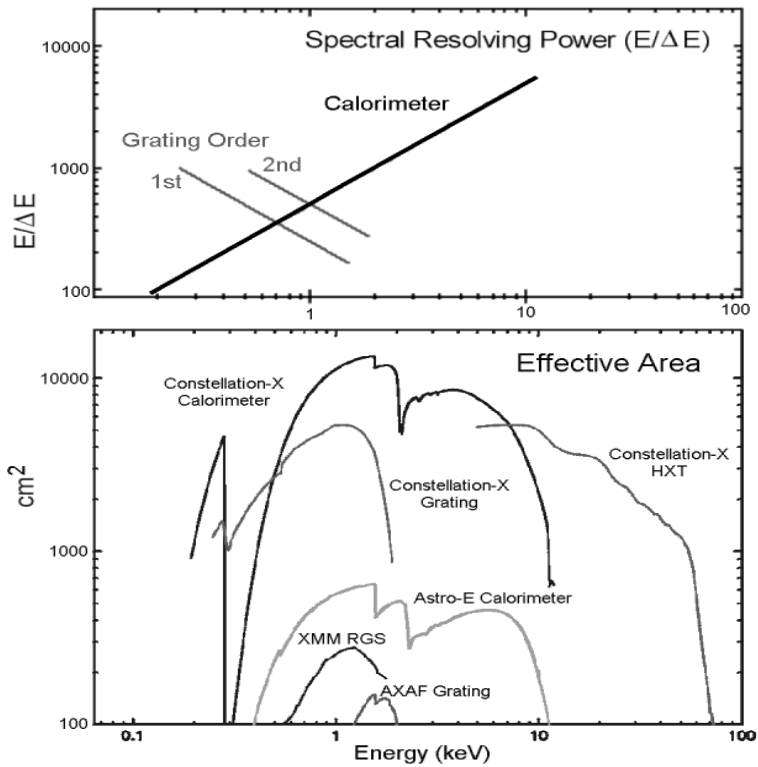,width=2.0in,height=3.2in}
\caption{
Spectral resolving power and the effective area of the instruments
onboard Constellation-X. 
}\label{myfirstfigure}
\end{figure}

The HXT uses a multilayer coatings on individual mirror shells to provide
the first focusing optics system to operate in the 6-40 keV band. Compared
to other non-focusing methods such as those used for RXTE, Constellation-X
has twice the area, 640 times the energy resolution, 240 times the
spatial resolution, and above 10 keV, 100 times the sensitivity. AXAF and
XMM, designated as the workhorses of X-ray astronomy in the next decade,
will detect photons with energies up to ~ 10 keV.

The technology development program is now underway and is targeting
a first launch in 2007-2008, around the time that AXAF will be reaching
the end of its projected lifetime. An essential feature of the Constellation-X
concept involves minimizing cost and risk by building several identical,
modest satellites to achieve a large area. The current baseline is 6 
satellites, although  other multiple satellite configurations are also
under consideration, with the final choice to be made based on 
a balance of overall cost and risk. 
The mission will be placed into a high
earth or L2 orbit to facilitate high observing efficiency, provide
an environment optimal for cryogenic cooling, and simplify the spacecraft
design. 

\section*{Science Goals}

Constellation-X is a key element in NASA's Structure and Evolution of the
Universe (SEU) theme aimed at understanding the extremes of gravity and
the evolution of the Universe. We highlight here a few key science areas.

{\bf How can we use observations of black holes to test
General Relativity?}
X-ray observations directly probe physical conditions close to the central
engine of blackholes where the distortions of time and space predicted
by general relativity are most pronounced. Constellation-X will use the
spectral features of these objects (e.g. the broad iron K$\alpha$ line 
discovered by ASCA \cite{tan95}) to 
map out the geometry of the inner emission regions and determine the 
extent to which we can test general relativity.  

{\bf What is the total energy output of the Universe?}
Models of cosmic X-ray background predict that the emission at hard
X-rays is due to many absorbed AGN \cite{mad94}, with their  
central engines primarily visible via hard X-rays (and perhaps infrared). 
If most of the accretion in the Universe is highly obscured, then the
emitted power per galaxy based on currently available optical, UV, or
soft X-ray quasar luminosity functions may be substantially underestimated.
By using hard X-ray spectra to advance our knowledge of the total
luminosity of AGN, Constellation-X will bring us closer to knowing the 
total energy output of the Universe.  
 
{\bf What roles do supermassive black holes play in galaxy
evolution?}
Constellation-X measurements of black hole mass and spin
for the high z quasar sample will allow understanding of the 
relative evolution rates of black holes and their host galaxies, and
will shed light on when massive black holes formed compared to the
galaxy formation epoch. 

{\bf How does gas flow in accretion disks and how do cosmic
jets form?}
Accretion disks play a fundamental role in many astrophysical settings,
ranging from the formation of planetary systems to accretion onto 
supermassive black holes in AGN. There are, however, many controversies
about the nature of viscosity which drives the accretion process,
about the stability of the disk at various accretion rates, and 
about the relevance of
advection and mass outflows, and the mechanisms by which jets are formed.
Constellation-X will probe the physics of accretion disks to a level
not currently possible, by resolving line features from the accretion 
disk photosphere and by measuring the continuum shape over a 
broad energy band.

{\bf When were clusters of galaxies formed and how do they
evolve?}
To date, cluster 
abundances have been measured in the X-ray band out to a redshift of
about 0.4 but no discernible evolution with z has been seen. 
Constellation-X spectra
of clusters over a range of redshifts will provide crucial information
about the presence of primordial gas, including any input from possible 
pre-galactic generations of stars as well as the contribution from 
stellar nucleosynthesis as a function of time. The high sensitivity
of Constellation-X is essential for extending such studies to the ``poorer
cousins'' of clusters, groups of galaxies. Moreover, by mapping
the velocity distribution of hot cluster gas via Doppler shifts in the
emission lines, Constellation-X will allow us to examine 
the effects of collisions
and mergers between member galaxies and between separate subclusters and
clusters. 

{\bf Where are the ``missing baryons'' in the local Universe?}
Recent observations of the Lyman-$\alpha$ forest show that at large
redshifts most of the predicted baryon content of the Universe is in the
IGM, while at low redshifts, the baryon content of stars, neutral hydrogen,
and X-ray emitting cluster gas is roughly one order of magnitude smaller
than that expected from nucleosynthesis arguments. Therefore, a large
fraction of baryonic content of the local Universe is considered ``missing''.
Numerical simulations \cite{cen98} predict that the missing
matter may reside in the IGM with a temperature range of $10^5-10^7$~K.
Such gas in the IGM can be detected with the high sensitivity, high resolution
instruments aboard Constellation-X through the absorption lines of
metals against the X-ray spectra of background quasars (e.g. OVII and OVIII).

{\bf How are matter and energy exchanged between stars and the Interstellar
Medium and how is the Intergalactic Medium enriched?}
The chemical enrichment of the Universe is dominated by star formation 
and the release of the processed material into the ISM via stellar
winds and supernova explosions. Moreover, supernova explosions
and enhanced star forming activities can drive hot gas out of the galaxy and
enrich the ICM/IGM on megaparsec scales. Detailed, spatially-resolved 
X-ray spectra reveal the stellar/supernova abundances, the composition of the 
surrounding ISM, and the interaction of the expanding blast wave with
the surrounding material. High throughput instruments such as those aboard
Constellation-X are needed to measure the K-lines of less abundant 
elements such as
F, Na, Al, P, Cl, K, Sc, Ti, V, Cr, Mn, Co, Ni, Cu, and Zn. The increased
sensitivity of Constellation-X will allow us to extend these studies
to exernal galaxies, beyond the Magellanic Clouds to M1 and M33, for example.
This will allow us to further our understanding of the history of star 
formation and exchange of matter between the ISM and stars.

\end{document}